\documentclass[preprint,amsmath,amssymb,aps]{revtex4-1}
\usepackage{graphicx}
\usepackage{amsfonts}
\usepackage{bm}
\usepackage{color}

\begin{document}

\newcommand{\G}{{\cal G}}
\renewcommand{\r}{{\rho}}
\newcommand{\cv}{{\bm c}}
\newcommand{\xv}{{\bm x}}
\newcommand{\s}{{\sum}}

\newcommand{\uv}{{\bm u}}
\newcommand{\qv}{{\bm q}}
\newcommand{\Fv}{{\bm F}}
\newcommand{\uvb}{\bar {\bm u}}
\newcommand{\te}{\theta}
\newcommand{\teb}{\bar{\theta}}

\newcommand{\T}{T}
\newcommand{\uvh}{{\bm u}_H}
\newcommand{\teh}{\theta_H}
\newcommand{\half}{\frac{1}{2}}

\newcommand{\vv}{{\bm v}}
\newcommand{\wv}{{\bm w}}
\newcommand{\dt}{{\Delta t}}
\newcommand{\Dt}{{D_t}}
\newcommand{\pii}{{\partial_i}}

\newcommand{\pj}{{\partial_j}}
\newcommand{\pk}{{\partial_k}}
\newcommand{\ppt}{{\partial_t}}
\newcommand{\f}{{f^{(0)}}}
\newcommand{\ff}{{f^{(1)}}}
\newcommand{\cs}{c_s^2}
\newcommand{\dij}{\delta_{ij}}
\newcommand{\dil}{\delta_{il}}
\newcommand{\dik}{\delta_{ik}}
\newcommand{\djk}{\delta_{jk}}
\newcommand{\djl}{\delta_{jl}}
\newcommand{\dkl}{\delta_{kl}}

\newcommand{\intv}{{\int dv^3}}

\title{A new assessment of the second order moment of Lagrangian
  velocity increments in turbulence}

\author{A.S. Lanotte$^{\rm a,e}$ $^{\ast}$\thanks{$^\ast$Corresponding
    author. Email: a.lanotte@isac.cnr.it\vspace{6pt}},
  L. Biferale$^{\rm b,e}$, G. Boffetta$^{\rm c,e}$, and F. Toschi$^{\rm d,e}$\\
\vspace{6pt} $^{\rm a}${\em{CNR-ISAC and INFN, Sez. Lecce,
    Str. Prov. Lecce-Monteroni, 73100 Lecce, Italy}}\\$^{\rm
  b}${\em{Dept. Physics and INFN, University of Rome ``Tor Vergata'',
    Via della Ricerca Scientifica 1, 00133 Rome, Italy}}\\ $^{\rm
  c}${\em{Dept. Physics and INFN, University of Torino, via
    P.Giuria 1, 10125 Torino, Italy}}\\$^{\rm d}${\em{Department of
    Applied Physics and Department of Mathematics \& Computer Science,
    Eindhoven University of Technology, Eindhoven, 5600MB, The
    Netherlands\\and CNR-IAC, Via dei Taurini 19, 00185 Rome, Italy}}\\
$^{\rm e}${Kavli Institute for Theoretical Physics China, CAS, Beijing 100190, China}}

\begin{abstract}
The behavior of the second-order Lagrangian structure functions on
state-of-the-art numerical data both in two and three dimensions is
studied. On the basis of a phenomenological connection between
Eulerian space-fluctuations and the Lagrangian time-fluctuations, it
is possible to rephrase the Kolmogorov $4/5$-law into a relation
predicting the linear (in time) scaling for the second order
Lagrangian structure function. When such a function is directly
observed on current experimental or numerical data, it does not
clearly display a scaling regime. A parameterization of the Lagrangian
structure functions based on Batchelor model is introduced and tested
on data for $3d$ turbulence, and for $2d$ turbulence in the inverse
cascade regime. Such parameterization supports the idea, previously
suggested, that both Eulerian and Lagrangian data are consistent with
a linear scaling plus finite-Reynolds number effects affecting the
small- and large-time scales. When large-time saturation effects are
properly accounted for, compensated plots show a detectable plateau
already at the available Reynolds number. Furthermore, this
parameterization allows us to make quantitative predictions on the
Reynolds number value for which Lagrangian structure functions are
expected to display a scaling region. Finally, we show that this is
also sufficient to predict the anomalous dependency of the normalized
root mean squared acceleration as a function of the Reynolds number,
without fitting parameters.

\end{abstract}

\maketitle

\section{Introduction}
The knowledge of the statistical properties of turbulence, and in
particular its non-Gaussian statistics, is a key open problem in
classical physics with important consequences for applications
\cite{frisch}. The description of a fluid flow can be equally done in
the Eulerian frame, where the velocity field at any position and time
is known, ${\bm u}({\bm x},t)$, or in the Lagrangian frame where the
evolution of fluid tracers, ${\bm x}(t)$, is followed in time, ${\bm
  v}(t) = {\bm u}({\bm x}(t),t)$ and ${\bm v}(t) = {\dot {\bm
    x}}(t)$. Although the two descriptions are mathematically
equivalent, the second bears premises to better shed light into the
dynamics of (small) particles dispersed and transported by turbulent
flows \cite{yeung,tb09}.

One of very few exact results known for three dimensional homogeneous
and isotropic turbulence is the Kolmogorov $4/5$-law for inertial
range of scales; for $d-dimensional$ flows with $d=2,3$, it reads as :
\begin{equation}
  \label{fourthird}
  S_3(r) = \left\langle [\left( {\bm u}({\bm x}+{\bm r}) - 
{\bm u}({\bm x})\right) \cdot \hat {\bm r}]^3\right\rangle = 
-\frac{12}{d(d+2)}\, \varepsilon  r\,,
\end{equation}
where longitudinal velocity increments are considered.

 This relation connects velocity differences at scale $r$ with the
 presence of a non vanishing energy flux, $\varepsilon$. In the $3d$
 direct cascade, the energy flux remains constant and positive at
 increasing the Reynolds number, giving rise to the dissipative
 anomaly of turbulence \cite{frisch}. The translation of
 Eqn. \ref{fourthird} to the Lagrangian domain has been suggested long
 time back \cite{Borgas,bof02} but it only relies on phenomenological
 bases. It connects Eulerian fluctuations at separation $r$, $\delta_r
 u = u(x+r)-u(x)$, with Lagrangian temporal velocity difference over a
 time interval $\tau$, $\delta_\tau v = v(t+\tau)-v(t)$, where space
 and time are connected through the local eddy turnover time:
\begin{equation}
\label{bridge}
\delta_\tau v \sim \delta_r u, \qquad \tau \sim r/\delta_r u\,.
\end{equation}
Here due to the {\it dimensional} and phenomenological nature of the
relation, all geometrical and vectorial properties are
neglected. Moreover, it is important to stress that the symbol $\sim$
in equation (\ref{bridge}) is meant as {\it scale-as} in a pure {\it
  statistical} sense and not as a deterministic constraint holding
point-by-point, as sometimes suggested \cite{grauer}. It results that
the phenomenological equivalent of the exact law (\ref{fourthird}) in
the Lagrangian domain reads:
\begin{equation}
\label{one}
S_2(\tau)  \equiv \langle (\delta_\tau v)^2 \rangle \sim \varepsilon \tau\,,
\end{equation}
where the prefactor cannot be exactly controlled. Another important
difference with respect to (\ref{fourthird}) is that the sign of the
right hand side is also fixed, implying that (\ref{one}) cannot be
exact in principle because of the energy flux differently sign-defined in
$2d$ and in $3d$ turbulence.

This relation is intimately connected with the picture of the
Richardson cascade, built in terms of a superposition of eddies at
different scales and with different characteristic times (eddy turn
over times). The idea is to imagine that Lagrangian fluctuations,
$\delta_\tau v$, at a given time-scale, $\tau$, are dominated by those
Eulerian eddies, $\delta_r u$, which have a typical decorrelation time
(\ref{bridge}) of the order of the time lag, $\tau$. Indeed eddies at
smaller scales are much less intense, i.e. if $r' \ll r$, then
$\delta_{r'} u \ll \delta_r u$, while eddies at larger scales do not
contribute to Lagrangian fluctuations being almost frozen on the time
lag $\tau$. The {\it bridge} relation (\ref{bridge}) must be
considered the zero-{\it th} order approximation connecting Lagrangian
and Eulerian domains. It cannot be exact and it cannot be applied
straightforwardly to all hydrodynamical systems, being strongly based
on the hypotheses of locality of the energy transfer process and on
the existence of a unique typical eddy turn over time at each
scale. Therefore it is not expected that it can straightforwardly
explain Lagrangian-Eulerian correlations in conducting flows, as
investigated in \cite{grauer_mhd}.\\ In considering the application of
the bridge relation for Lagrangian scaling in $2d$ and $3d$
hydrodynamical turbulence, the situation is not at all yet clear. On
the one hand, it has been successfully used to predict the probability
density function of accelerations and the relative scaling between
Lagrangian structure functions \cite{PRL2004,JOT2005,PRL,JFM}. On the
other hand, when looking at direct scaling {\it versus} the time lag,
inconclusive results have been obtained \cite{BY04a,mordant,LA02}.  As
a consequence, different scaling behaviors have been proposed to
overcome doubts raised due to the consistently poorer quality of the
validation from both numerical and experimental tests \cite{noi,loro},
when compared to the Eulerian counterpart. Moreover, by means of a
stochastic model, it has been argued that the observed reduced scale
separation in the Lagrangian frame is the main reason for the
departure from Kolmogorov scaling in data\cite{sawford}, and that the
inertial sub-range linear scaling is eventually reached only at
Reynolds numbers beyond $Re_{\lambda} = 30000$ \cite{SY}. \\ Note that
acceleration probability and relative scaling of Lagrangian structures
functions are a probe of intermittent fluctuations over time lags
$\tau$, which can be assessed independently of the scaling of the
second order moment $S_2(\tau)$. However, this deserves a particular
interest since it is a key ingredient of Lagrangian stochastic models
for turbulent diffusion and dispersion
\cite{yeung,Th90,Sawford2001,BY2004}. \\In this manuscript we
specifically address the issue of poor inertial range scaling of
$S_2(\tau)$, basing our analysis on currently available numerical
data.  Our analysis points in the direction of an enhanced sensitivity
to finite-Re corrections in the Lagrangian framework with respect to
the Eulerian one.  We show that a simple modeling of finite Reynolds
number effect, affecting the small and the large scales, can be enough
to interpret present data on the basis of the {\it bridge} relation
(\ref{bridge}). This result confirms two things.  First, that the
dimensional {\it Kolmogorov-like} argument of (\ref{bridge}), even if
not supported by any exact theoretical statement, represents a very
good first start to guess the statistical connections between Eulerian
and Lagrangian statistics.  Second, that any possible {\it new
  physics} beyond the relation (\ref{bridge}) is to be compared with
more refined data at higher Reynolds numbers.

We remark that on the basis of the refined similarity approach
\cite{RKSH,RKSH2}, no intermittency correction of the second order
structure function is expected. Alternatively, in \cite{loro}, a small
modification of the linear scaling for the second order structure
function has been proposed on the basis of the observed behavior of
the acceleration spectrum. While further data are needed to definitely
discriminate between anomalous scaling or finite Reynolds number effects in
the second order moment, we point out that the simple parameterization
here proposed gives very good results without invoking any
intermittent correction. Finally, we stress that recently, by using
Hilbert-Huang Transform, further evidences for a linear scaling of
second order Lagrangian moment have been presented\cite{bife_cina}.
\section{Batchelor parameterization for the Lagrangian second order
  structure function}
We consider the second order moment of velocity increments measured
along tracer trajectories in statistically stationary, isotropic and
homogeneous (HIT) $3d$ dimensional turbulence:
\begin{equation}
S_2(\tau)\equiv \langle [v_i(t+\tau) - v_i(t)]^2 \rangle\,,
\end{equation}
where $v_i(t)$ is one component of the turbulent Lagrangian velocity
field. As mentioned, the Kolmogorov scaling for the Eulerian velocity
increments once translated into the time domain via the {\it bridge}
relation gives -for any velocity component- the linear prediction
$S_2(\tau)= C_0\, \varepsilon \tau$ where $C_0$ is a dimensionless
constant of order unity.  Observations \cite{YPS2006} suggest that in
$3d$ HIT, $C_0 \in [6-7]$\,: however, since even at the largest
Reynolds number achieved, both experimental and numerical data do not
show a well developed scaling range in $S_2(\tau)$, the value of $C_0$
measured displays a weak yet detectable $Re$ dependence
\cite{YPS2006,noi,SY,loro,pinton,LA02}.

The point that we address here is to understand if this poor scaling
reflects a real deviation from the linear scaling of Lagrangian
turbulence, or if it is just the result of finite Reynolds numbers
effects, coming from both ultraviolet (UV) and infrared (IR)
cutoffs. In the latter case, one could expect that future DNS and
experiments might be able to directly display scaling properties also
in the Lagrangian domain, including intermittency. In fact, at the
moment current practice is analysing intermittency in the Lagrangian
domain only by using Extended Self-Similarity approach \cite{PRL,JFM},
hence bypassing the need for well defined power-law behavior in the
inertial range.

In order to understand the above issue, it is mandatory to have a
control on the effects of viscous and integral scales on the {\it
  supposed} inertial range.  Due to the lack of control on the
analytical side, one possible way is to resort to phenomenological
models \cite{meneveau,sawford,PRL}, trying to reproduce the behavior
of the velocity increments over the entire range of
scale/frequency. In particular, according to studies over the last few
decades \cite{meneveau,PRL,sreeni,RT,JFM}, a parameterization proposed
by Batchelor became quite popular because of its simplicity and
capability to include non trivial viscous-effects (such as the
intermittency of velocity gradients and acceleration)
\cite{meneveau,PRL}, as well as the saturation effects observed at the
large scales \cite{sreeni,RT,JFM}.

In the following, we test the possibility to get a suitable {\it
  Batchelor}-like parameterization able to capture the poor scaling
behavior observed on the data. The anticipated success of this goal
implies two facts. First, it shows that the absence of a genuine
scaling observed at moderate Reynolds numbers {\it is not in
  contradiction} with the possibility to have scaling at higher
Reynolds numbers. Second, it gives a first hint on how far in Reynolds
number one needs to go before expecting an observable scaling
behavior.  Of course the Batchelor parameterization is not based on
any analytical result and finds a justification only on its ability to
reproduce data.  Other parameterizations are very much possible as
well, and whether the Batchelor one will agree or not with data at
higher Reynolds numbers is an open question for the future.

\begin{figure}
\begin{center}
  \includegraphics[width=9cm]{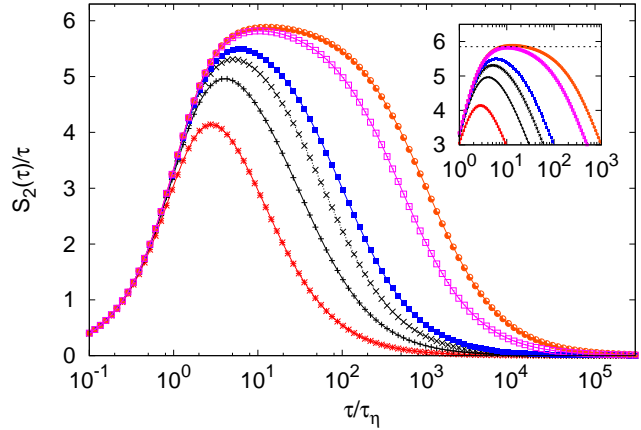}
  \caption{The linearly compensated second order Lagrangian structure
    function as obtained with the Batchelor parameterization
    (\ref{eq:param}), for different $Re_\lambda$. Starting from bottom
    curve, they refer to structure functions at the following values
    of Taylor-scale based Reynolds numbers $Re_λ=100; 300; 600; 1000;
    5000$ and $Re_{\lambda}=10000$.  The inertial range scaling
    exponent is fixed to $z_2=1$. Inset: a zoom in the scaling region
    to highlight a plateau starting to develop already at
    $Re_{\lambda}=5000$.}
\label{fig:fit}
\end{center}
\end{figure}

On a dimensional ground, a parameterization for the time behavior of
$S_2(\tau)$ has to reproduce the three following regimes:
\begin{equation}
\begin{cases}
  S_2(\tau ) \sim \epsilon\, \tau^2/\tau_{\eta} \qquad \tau \ll \tau_{\eta}\,,\\ S_2(\tau )
  \sim \epsilon\, \tau (\tau/T_L)^{z_2 -1} \qquad \tau_{\eta} \ll \tau
  \ll T_L\,,\\ S_2(\tau ) \sim \epsilon\, T_L  \qquad \tau \gg T_L\,,\\
\end{cases}
\label{eq:bachelor}
\end{equation}
where $\tau_{\eta}$ is the Kolmogorov time scale and $T_L$ is the
large scale Lagrangian eddy turnover time. If we assume a Kolmogorov
scaling in the temporal inertial range then $z_2=1$, otherwise it can
be kept as a free parameter (see also Sec. \ref{sec:acc}).  We recall
that by dimensional arguments we have $T_L/\tau_{\eta} \propto
Re_{\lambda}$. A functional form which interpolates between the above
behaviors is simply obtained as \cite{meneveau,sreeni,RT}:
\begin{equation} 
S_2(\tau) = C_0\,\epsilon\, \frac{\tau^2}{(c_1\,\tau_{\eta}^2 + \tau^2)^{\frac{(2-z_2)}{2}}}\,(1 + c_3 \tau/T_L)^{-z_2}\,,
\label{eq:param}
\end{equation} 
where $c_1$ and $c_3$ are order one dimensionless constants.\\In
Figure~\ref{fig:fit}, we show the results for the linearly compensated
second order moment, when we take $T_L/\tau_{\eta} = 0.1 Re_{\lambda}$
\cite{YPS2006}. It turns out that the effect of finite Reynolds number
induced by the large scale saturation are big, since a plateau
develops only for very large Reynolds numbers currently
unreachable. In the inset, we zoom in the scaling region: starting for
Reynolds number $Re_{\lambda} = 5000$, a scaling shows up. \\ One can
of course play with the parameterization in order to modify the
transitions from viscous to inertial, and from inertial to integral
ranges. In particular, by changing the functional form of the
denominator in eqn. (\ref{eq:param}) and of the saturation factor,
these transitions can be made sharper or smoother \cite{PRL}. \\ We
also note that in order to be consistent with an exponential decay for
the velocity correlation function, one can possibly slightly refine
the functional form of the saturation factor for large times (see
below). It is thus probable that the observed absence of a clear and
well developed plateau in numerical and experimental data is just a
finite Reynolds number effect that, as we mentioned, are more
pronounced in the Lagrangian statistics than in the Eulerian case
(dimensionally, $L/\eta \propto Re_{\lambda}^{3/2}$ and $T_L/\tau_\eta
\propto Re^{1/2}$).

\begin{figure}
  \includegraphics[scale=0.58]{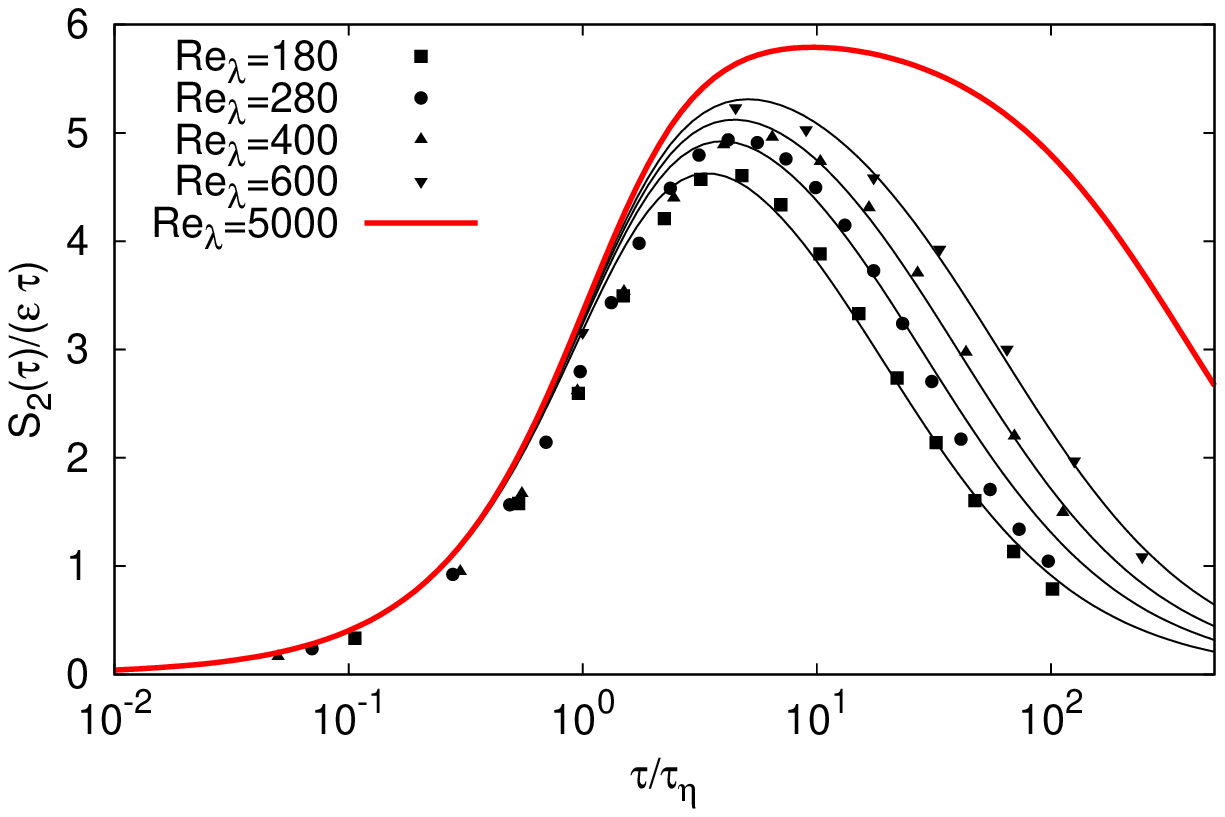}
  \includegraphics[scale=0.58]{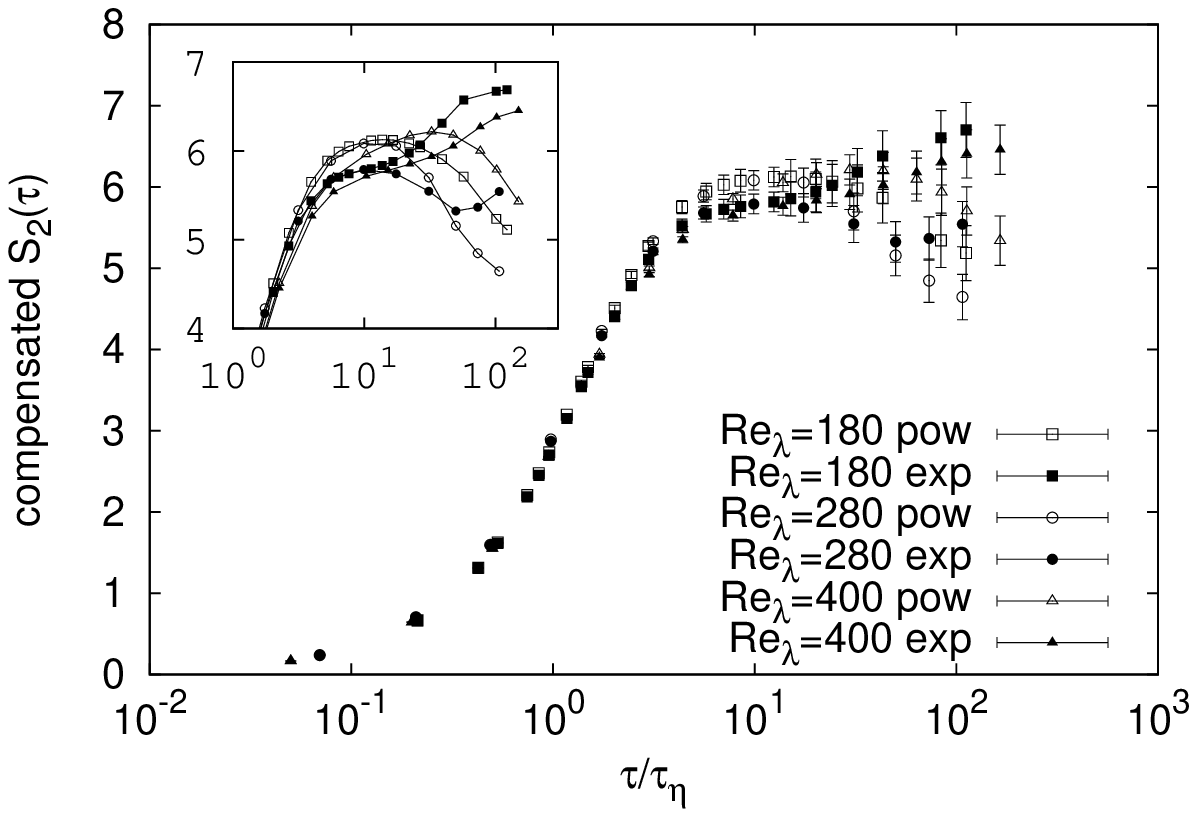}
  \caption{(Left panel) DNS of $3d$ HIT at $Re_{\lambda} \sim 180,
      280, 400, 600$ \cite{pof,JFM2,JFM}. $S_2(\tau)$ compensated with
      $\varepsilon \tau$ versus the Batchelor fit (solid lines) with a
      power-law large-scale saturation term. (Right panel) Same DNS
      data of HIT at three different Reynolds numbers, compensated
      such as to highlight inertial range behavior according to the
      two Batchelor parameterizations (with large times exponential or
      power-law behavior, see eqn. \ref{largescale}). Fitting
      parameters are: $c_1=2.2$ in the power-law and $c_1=2.5$ in the
      exponential form; $c_3=1.0$ in the power-law expression, while
      $c_3=1.5$ in the exponential expression; in all cases
      $C_0=6$. Error bars are estimated from the anisotropy of
      velocity components at large scales. (Inset of right panel) same
      curves as in the body, to highlight large scale behavior. Error
      bars have been omitted for clarity.}
    \label{fig:data}
  \end{figure}

In Fig.~(\ref{fig:data}), we present an analysis of DNS data of $3d$
HIT at $Re_{\lambda}=180, 280, 400, 600$ (see \cite{pof,JFM2,JFM}). In
particular, we compare the linearly compensated second order
Lagrangian structure functions at the four Reynolds numbers (left
panel), with curves obtained according to eqn. (\ref{eq:param}). As
one can see the fit is very good.  Moreover, in the same figure we
also show, to guide the eyes, the result of the Batchelor
parameterization for a much higher Reynolds number ($Re_\lambda =
5000$). \\
It is well known that time correlations along tracer trajectories
decay very slowly. Hence, when considering the second order Lagrangian
structure function, there is the issue of the long time decaying of
the velocity correlation functions. Here we compare the power-law
saturation factor $\propto (1+ \tau/T_L)^{-z_2}$ appearing in
eqn.~(\ref{eq:param}), with an exponential saturation factor ruling
the large times behavior. We used the following interpolation\,:
\begin{equation} 
  S^{*}_2(\tau) = C_0\, \epsilon\,T_L\, \frac{\tau}{(c_1\tau_{\eta}^2 +
    \tau^2)^{1/2}} (1- \exp(-c_3\tau/T_L))\,,
\label{eq:inter}
\end{equation} 
where in comparison to expression (\ref{eq:param}), we have fixed the
exponent $z_2=1$ and $C_0$, $c_1$ and $c_3$ are free parameters. In
the right panel of Fig.~(\ref{fig:data}), we compare the results of
the two different functional forms for large time scales. In order to
do it properly, we plot the second order structure function
compensated with its whole inertial and integral time scale regime, that is
\begin{equation}
\frac{S_2(\tau)}{ \left(\tau \,(1 + c_3 \tau/T_L)^{-1}\right)}; \qquad  \frac{S^*_2(\tau)}{(\tau \,(1- \exp(-c_3\tau/T_L)))}\,,
\label{largescale} 
\end{equation}
though the power-law and exponential forms are very close. Here
clearly it is important to consider that at large scales we expect to
have quite {\it large} statistical and systematic error bars, due to
anisotropy and/or finite size effects (see error bars in the right
panel of Fig.~\ref{fig:data}). Moreover, large scale fluctuations are
not expected to be universal. It is interesting to note, that once the
large scale contamination is removed, compensated data start to show a
well-defined plateau already at mode\-ra\-te Reynolds numbers,
independently of the functional form for the large scale
behavior.\\ 
\begin{figure}
\centering
  \includegraphics[width=9cm]{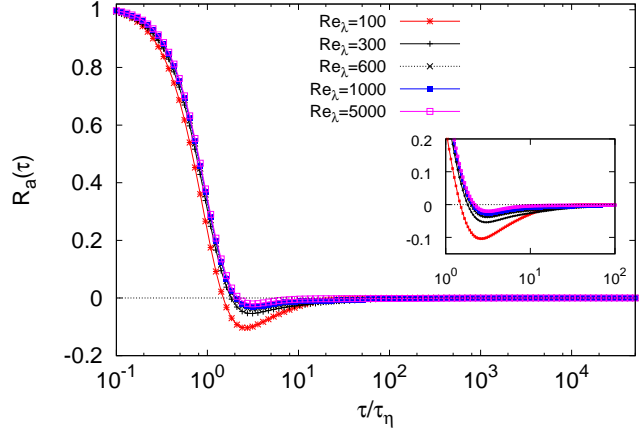}
  \caption{The temporal behaviour of acceleration autocorrelation
    function $R_a(\tau)= \langle a(\tau) a(0)\rangle$ derived from the
    Batchelor parameterization of the Lagrangian second order
    structure function, with exponential large-scale saturation, see
    eq.~(\ref{eq:inter}). Parameters are the same used for
    Figure~\ref{fig:data}. In the inset, a zoom in the negative region
    of the autocorrelation function.}
    \label{fig:acov}
  \end{figure}
Scaling relations have to be consistent with kinematic
constraints of - statistically stationary and isotropic-,
turbulence. One of such is that the time integral of the acceleration
autocorrelation function is zero \cite{TL72,BS91}\,,
\begin{equation}
\int_0^{\infty}R_a(\tau) d\tau =\int_0^{\infty} \langle a_i(\tau)
a_i(0)\rangle d\tau=0\,.
\label{eq:acov}
\end{equation}
Hence the acceleration autocorrelation, which is positive at small
time lags, should then be negative to match the kinematic
constraint.\\ Provided a linear leading scaling is prescribed in the
Lagrangian second order structure function, the acceleration
autocorrelation function is further constrained to be zero in the
inertial range of scales \cite{BS91}. In Fig.~\ref{fig:acov}, we plot
the behaviour of the acceleration covariance obtained from the
Batchelor parameterization of the Lagrangian second order structure
function, which shows consistency vanishing behaviour in the scaling
range. These findings are valid as possible working hypothesis, until
as suggested in \cite{BS91} and \cite{loro}, a precise form of the
acceleration autocorrelation is known both in the dissipative and
inertial subranges for finite Reynolds number 3D turbulence.\\ Hence,
at least for $3d$ turbulence, we summarize these indications as
follows: (i) the absence of a plateau can be related to the presence
of strong large-scale and small-scale effects, competing with the
inertial range behavior; (ii) as it appears from the left and the
right panels of Fig.~\ref{fig:data}, the Lagrangian inertial range
does not coincide with the plateau region, where the second order
structure function linearly compensated shows a peak, since the
large-scale contamination is still present.

\section{Intermittency corrections}
\label{sec:acc}
\begin{figure}
\begin{center}
  \includegraphics[width=9cm]{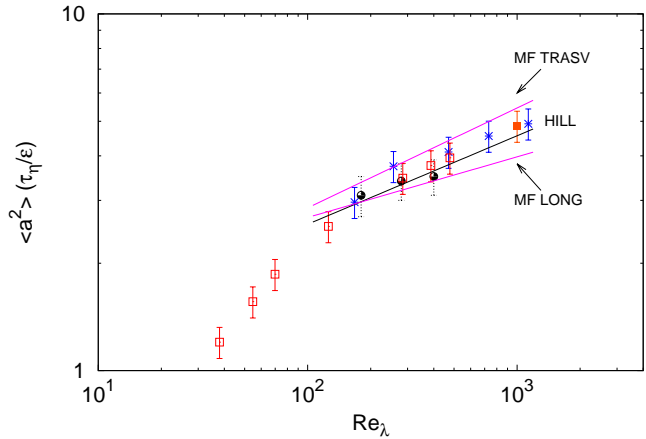}
  \caption{Collection of different numerical data of the scaling of
    normalized root-mean-square acceleration as a function of the
    Taylor-scale based Reynolds number $Re_\lambda$. Two lines
    correspond to the multifractal prediction using the bridge
    relation for transverse increments (MF TRASV) leading to $\gamma =
    0.17$, or the bridge relation for longitudinal increments (MF
    LONG) leading to $\gamma=0.28$ (see \cite{PRL} for details). These
    lines can be shifted up or down arbitrarily, being the
    multifractal prediction valid scaling-wise and not for the
    prefactors. A third line is a fit proposed by R. Hill in
    \cite{Hill}, as a superposition of two power laws of exponents
    $\gamma_1=0.25$ and $\gamma_2=0.11$.  Data are taken from
    Refs. \cite{pof,JFM,JFM2,vedula,gotoh,kaneda}.  Error bars are
    estimated considering a typical $10\%$ uncertainty in the energy
    dissipation rate.}
\label{fig:acc}
\end{center}
\end{figure}

\begin{figure}
\begin{center}
  \includegraphics[width=9cm]{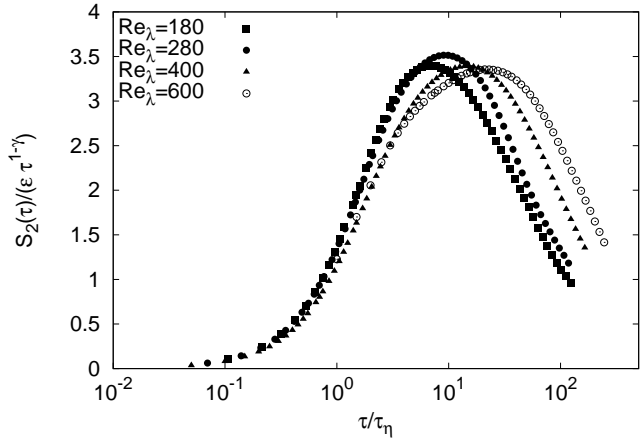}
  \caption{The second order Lagrangian structure function compensated
    as $S_2(\tau)/(\epsilon \tau^{1-\gamma})$, with $\gamma = 0.22$. The
    anomalous correction $\gamma$ is extracted from acceleration data
    shown in Fig.~(\ref{fig:acc}).
    \label{fig:newcomp}}
\end{center}
\end{figure}

It is well known that Lagrangian statistics in $3d$ is affected by 
intermittent corrections. In particular, acceleration statistics does
not obey dimensional scaling: the normalized
acceleration rms, $\langle a^2 \rangle \tau_\eta/\varepsilon$ is
observed to have a {\it weak} anomalous dependency on $Re_\lambda$:
\begin{equation}
  a^2_{rms}= \langle a^2 \rangle\,  \sim a_0\, \frac{\varepsilon}{\tau_\eta} \,Re_\lambda^{\gamma}\,,
\label{eq:a}
\end{equation}
with $\gamma \sim 0.2$ (see also Fig.~\ref{fig:acc}). Similarly, the
probability density function of the normalized acceleration
$P(a/a_{rms})$ possesses strong non-Gaussian and Reynolds-dependent
tails \cite{PRL2004}. It is remarkable that such intermittent
corrections can be explained by invoking again the {\it bridge }
relation previously discussed: so doing, it is possible to predict
Lagrangian intermittent properties once the Eulerian ones are given,
and viceversa \cite{bof02,PRL2004,PRL,pof,JFM}. In
Fig.~(\ref{fig:acc}), we report the compilation of data sets at
different Reynolds numbers for the root-mean-square acceleration,
(\ref{eq:a}). On these data three curves are superposed\,: (i) a
phenomenological fit proposed in \cite{Hill}, and the predictions
obtained by using the bridge relation (\ref{bridge}) with two
different multifractal estimates of the Eulerian statistics, based on
the longitudinal and on the transverse spatial increments
\cite{JFM}. The numerical data fall well within the two multifractal
predictions, confirming the ability of the bridge relation to
reproduce Lagrangian properties {\it without } any additional free
parameter. We notice that the bridge relation still predicts that
(\ref{one}) holds true, i.e.  intermittency is absent for third-order
quantities in the Eulerian domain and {\it hence} for second order
quantities in the Lagrangian one.\\ An alternative approach can be
followed by assuming {\it independent} anomalous scaling properties
for Lagrangian and Eulerian domains, i.e. without using the bridge
relation.  In this case, $S_2(\tau)$ is not constrained to scale
linearly and one could assume a pure inertial range intermittent
correction as in eq.~(\ref{eq:bachelor}) with $z_2=1-\gamma$\,,
\begin{equation}
\label{eq:inte}
S_2(\tau) \sim \varepsilon \tau \left(\frac{\tau}{T_L}\right)^{-\gamma}\, \tau \gg \tau_{\eta}\,,
\end{equation}
where $\gamma$ is no longer linked to any Eulerian properties;
moreover, $\tau_{\eta}$ must fluctuate independently of the Eulerian
fields, too. This is another route to explain the anomalous scaling of
the acceleration variance as a function of $Re_\lambda$, which has
been investigated in \cite{loro} and which is not in contradiction
with any exact scaling law in Lagrangian turbulence. In \cite{loro},
the intermittent correction $\gamma$ was obtained from a fit of the
scaling of (\ref{eq:a}).\\
In Fig.~(\ref{fig:newcomp}) we apply the intermittent compensation
$\tau^{1-\gamma}$ to the DNS data shown in previous sections and
observe that the plateau is slightly increased, but it is still very
narrow. {\it Finite Reynolds number} effects are overwhelming.\\ The
question whether $S_2(\tau)$ scales linearly, or with an intermittent
correction, or does not scale at all, needs data at higher Reynolds
numbers to further support present ideas.

\section{Inverse cascade in $2d$ turbulence}
In this section, we present results on Lagrangian structure functions
measured in the inverse cascade regime of $2d$ homogeneous and
isotropic turbulence. Again, the general question we want to address
is whether Lagrangian statistics are compatible with Eulerian
statistics, i.e. if a suitable transformation from space to time is
able to reproduce Lagrangian statistics given the Eulerian one. We
remind ourselves that, in spite of the fact that the inverse cascade is
statistically {\it simpler} than the direct cascade in $3d$ (since in
$2d$ inverse cascade, the Eulerian statistics displays Kolmogorov
scaling without intermittency corrections \cite{BE12}), a recent work
\cite{kf08} claims that Lagrangian statistics do not reflect this
simplicity and cannot be related to Eulerian statistics.

In the following, we consider Eulerian and Lagrangian structure
functions obtained from numerical simulations of $2d$ Navier-Stokes
equations for the vorticity $\omega = \nabla_x u_y - \nabla_y u_x$:
\begin{equation} \frac{\partial \omega}{\partial t} + ({\bf u} \cdot
  {\bf \nabla}) \omega = \nu \nabla^{2} \omega - \alpha\, \omega +
  f_{\omega}\,,
\label{eq:2.1}
\end{equation}
in the inverse cascade regime at resolutions $2048^2$. The forcing
$f_{\omega}$ is active on a range of wavenumbers around $k_f \simeq
256$, is $\delta$-correlated in time and injects energy at a fixed
rate $\varepsilon_I$. About one half of the injected energy flows to
large scale generating the inverse cascade with a flux
$\varepsilon$. The $\alpha\, \omega$ friction term is necessary to
reach a stationary state, and defines the large-scale eddy turnover
time $T_L \simeq 1/\alpha$. Different runs correspond to different
values of $\alpha$ and therefore to different extension of the
inertial range of scales. The smallest characteristic time, the {\it
  Kolmogorov time} $\tau_{\eta}$, is given in the inverse cascade by
the time at the forcing scale $l_f \sim 1/k_f$, that is kept
fixed. The extension of the inertial range in the time domain is thus
growing as $T_L \propto 1/\alpha$.

To compare Eulerian and Lagrangian structure functions, a simple
model, motivated by the cascade model for turbulence, can be
introduced. We represent turbulent Eulerian velocity fluctuations
$\delta_r u$ as the superposition of the contributions from different
eddies in the cascade \cite{BBCCV_PRE98}\,: $\delta_r u = \sum_n u_n
f(r/r_n)$, where $u_n$ is the typical fluctuation at the scale
$r_n$. The decorrelation function $f(x)$ is such that $f(x) \sim x$ as
$x \ll 1$ and $f(x) \sim 1$ for $x \gg 1$\,: here, we choose the
simple function $f(x)= 1-exp(-x)$.\\Within this framework, it is
natural to represent the corresponding Lagrangian velocity fluctuation
as $\delta_{\tau} v = \sum_n v_n f(\tau/\tau_n)$ where $\tau_n \sim
r_n^{2/3}$ is the correlation time of the eddy at scale $r_n$. A
minimal realization of this model requires the presence of two scales
which govern the crossover from dissipative to inertial scales,
$\eta$, and from inertial to integral scales, $L$. We can therefore
write, introducing explicitly the scaling behavior in the inertial
range, the following relation
\begin{equation}
  \delta_r u = U\, f\left(\frac{r}{L} \right) + 
  U\, \left[1-f\left( \frac{r}{L} \right) \right] f\left( \frac{r}{\eta} \right)
  \left( \frac{r + \eta}{L} \right)^{1/3}\,,
\label{eq:2.2}
\end{equation}
which, for Lagrangian increments, translates into
\begin{equation}
  \delta v(\tau) = U_L f\left( \frac{\tau}{T_L} \right) + 
  U_L \left[1-f\left( \frac{\tau}{T_L} \right) \right] 
  f\left( \frac{\tau}{\tau_{\eta}} \right)
  \left( \frac{\tau + \tau_{\eta}}{T_L} \right)^{1/2}\,.
\label{eq:2.3}
\end{equation}
$U$ and $U_L$ are the root-mean-square velocities in the Eulerian and
Lagrangian domain, respectively.
\begin{figure}[h]
\centerline{\includegraphics[width=9cm]{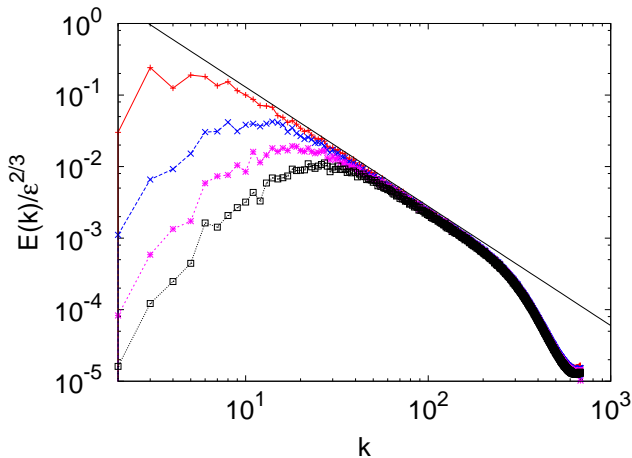}}
\caption{Kinetic energy spectra from $2d$ direct numerical simulations
  at resolution $N^2=2048^2$, with $\alpha=0.02$ (red $+$),
  $\alpha=0.04$ (blues $\times$), $\alpha=0.06$ (pink $*$) and
  $\alpha=0.08$ (black $\square$). The line represents Kolmogorov
  spectrum $E(k)=C \varepsilon^{2/3} k^{-5/3}$ with $C=6$.}
\label{fig2.1}
\end{figure}
In Fig.~\ref{fig2.1}, it is shown that in the stationary state, we
observe an inverse cascade with a Kolmogorov spectrum which extends
from the forcing wavenumber $k_f=256$ to the friction wavenumber
$k_{\alpha} \simeq \varepsilon^{-1/2} \alpha^{3/2}$ \cite{BM10,BE12}.\\
\begin{figure}[h]
  \centerline{\includegraphics[width=9cm]{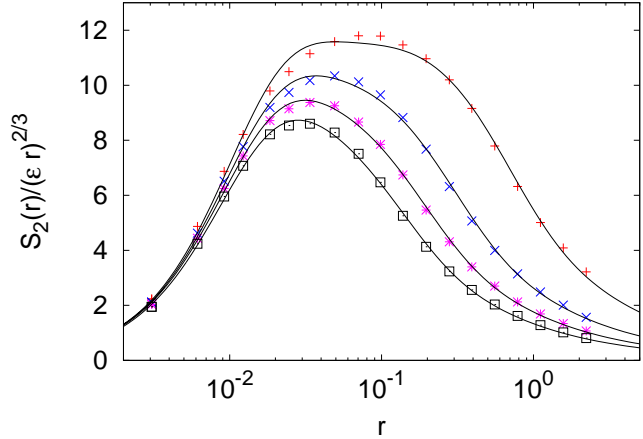}}
  \caption{Eulerian second-order structure function $S_2(r)$ in the
    inverse energy cascade, compensated with Kolmogorov scaling
    $(\varepsilon r)^{2/3}$. Colors and symbols as in
    Fig.~\ref{fig2.1}. Lines represent the fit with $(\delta_r u)^2$
    as in eq.~(\ref{eq:2.2}), which gives the ratio $L/\eta=12$
    ($\alpha=0.08$), $L/\eta=16$ ($\alpha=0.06$), $L/\eta=25$
    ($\alpha=0.04$) and $L/\eta=54$ ($\alpha=0.02$).}
\label{fig2.2}
\end{figure}
In Fig.~\ref{fig2.2} we show the Eulerian second-order structure
functions $S_2(r)=\langle (\delta_r u)^2\rangle$, compensated with
dimensional scaling $(\varepsilon r)^{2/3}$, for different values of
$\alpha$. An important remark is that, in spite of the clear power-law
scaling in the spectra, we do not observe any inertial range scaling
for the Eulerian structure functions, even for the most resolved
simulation. Nonetheless, the simple two-scales model, for the Eulerian
statistics (\ref{eq:2.2}) and for the Lagrangian one (\ref{eq:2.3}), is
able to reproduce quite accurately the crossovers from dissipative and
to integral scales, with parameters ($\eta/L,\tau_{\eta}/T_L,U,U_L$)
which change according to dimensional predictions (cfr. caption of
Figs.~\ref{fig2.2} and \ref{fig2.3}).
\begin{figure}[h]
  \centerline{\includegraphics[width=9cm]{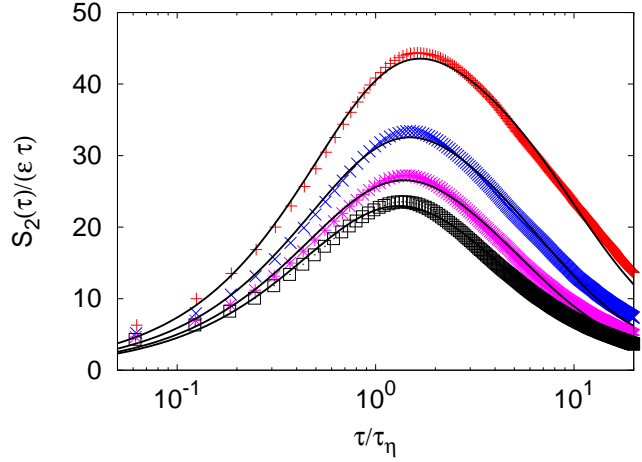}}
  \caption{Second order Lagrangian structure function $S_2(\tau)$ in
    the inverse energy cascade, compensated with $\varepsilon
    \tau$. Colors and symbols as in Fig~\ref{fig2.1}. Lines represent
    the fit with $(\delta_{\tau} v)^2$ as in eq.~(\ref{eq:2.3}), which
    gives the ratio of times $T_L/\tau_{\eta}=7.8$ ($\alpha=0.08)$,
    $T_L/\tau_{\eta}=8.9$ ($\alpha=0.06)$, $T_L/\tau_{\eta}=11.2$
    ($\alpha=0.04)$ and $T_L/\tau_{\eta}=16.1$ ($\alpha=0.02)$.}
  \label{fig2.3}
\end{figure}
Lagrangian structure functions $S_2(\tau)$ linearly compensated with
$\varepsilon \tau$ are shown in Fig.~\ref{fig2.3}, together with the
prediction obtained from model (\ref{eq:2.3}). The model fits well the
data, at least at small and intermediate times and with parameters
which change with the extension of the inertial range.\\ We remark
that the model parameterization - as well as the Batchelor model or any
other model, see \cite{sawford,chevillard}- are all constructed on the
hypothesis of a linear scaling in the inertial sub-range. The point we
want to make here is that, within the approximation model, the quality
of data fit is comparable for Eulerian and Lagrangian
statistics.\\More sophisticated multi-scale models can be envisaged,
e.g. based on the superposition of a hierarchy of characteristic
scales and times, at the price of a complex form of the
parameterization.
\begin{figure}[h]
\centerline{\includegraphics[width=9cm]{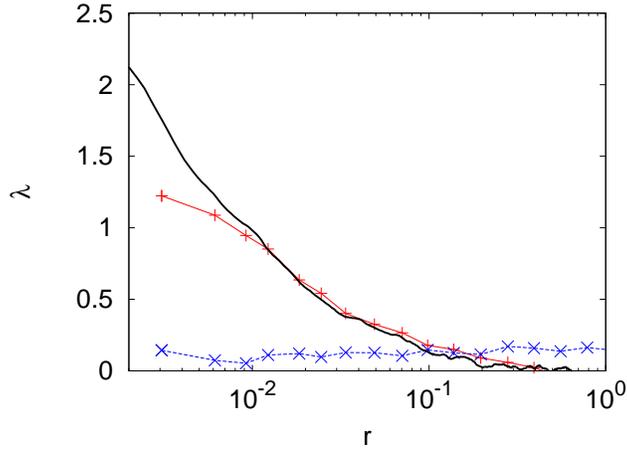}}
\caption{Excess kurtosis $\lambda(r)=S_4/S_2^2-3$ for Eulerian
  longitudinal structure function ($\times$ blue); for the Eulerian
  $x$-component structure function ($+$ red) averaged over the
  increment vector ${\bf r}$ taken in all directions; for the
  Lagrangian structure function (black line) with time rescaled as
  $r=0.035 \tau^{3/2}$.  Data refer to the run with $\alpha=0.02$.}
\label{fig2.4}
\end{figure}

One interesting result discussed in \cite{kf08} is that Lagrangian
statistics in two dimensional turbulence are not Gaussian, even if
Eulerian statistics are very close to Gaussian in the inverse cascade.
Our simulations confirm this result but suggest that this is a
delicate point as the statistics may depend on the observable.
Figure~\ref{fig2.4} shows the excess kurtosis for Lagrangian structure
function $\lambda(\tau)\,=\, \left[S_4(\tau)/S_2(\tau)^2-3\right]$ and
for Eulerian structure function $\lambda(r)\,=\,
\left[S_4(r)/S_2(r)^2-3\right]$, measured for both the $x$-component
velocity increments and for the longitudinal velocity increments. The
kurtosis of longitudinal velocity increments is constant and close to
Gaussian value at all scales, but this is not the case for Eulerian
increments of a single component of the velocity. We do not have a simple
explanation for this observation, but we think that this is a possible
origin of the deviation from Gaussianity observed in Lagrangian
statistics. Indeed, figure~\ref{fig2.4} also shows that the Lagrangian
excess kurtosis $\lambda(\tau)$ is very close to $\lambda(r)$, when
time is rescaled using the {\it bridge} relation $\tau_r=c
r^{2/3}$. Of course this rescaling can work only in the inertial range
and therefore we observe deviations at small separation $r$.

\section{Conclusions}
The Lagrangian and Eulerian description of the velocity field of a
fluid are of course correlated and it should be possible to rephrase
some statistical properties of Eulerian turbulence in terms of
Lagrangian counterparts and viceversa. The question is ``how much''
and ``what'' one can bridge by using relation (\ref{bridge}).\\
The simplest phenomenological description connects velocity
fluctuations in time to velocity fluctuation in space, $\delta_r u
\sim \delta_\tau v$, where the time-lag, $\tau$, and space separation,
$r$, are connected by the relation $\tau \sim r/\delta_r u$. From such
a connection, one can obtain the prediction that $S_2(\tau) \sim
\varepsilon \tau$, independently of the Eulerian intermittency, which
is the Lagrangian rephrasing of the Kolmogorov $4/5$-law.

As we discussed, both the linear scaling relation of $S_2(\tau)$, and
the Eulerian vs. Lagrangian mapping could be objected to. Reasons for
questioning their validity are: (i) the fact that such a relation,
contrarily to the $4/5$-law, is not rigorously derived; (ii) the fact
that the scaling of the $S_2(\tau)$ appears to be of poorer quality
than its Eulerian counterpart.\\ In the present manuscript we have
addressed the issue of the consistency of present state-of-the-art
numerical data with the linear dimensional scaling for the
$S_2(\tau)$, both in $3d$ and $2d$ turbulence. More specifically we
have tried to shed further light on the question whether or not the
present data are consistent with the linear scaling for the
$S_2(\tau)$ plus finite Reynolds number effects. Eulerian and Lagrangian
data, both for $3d$ and $2d$ turbulence, appear to agree equally well
with a Batchelor-like parameterization, which takes into account
dissipative and integral effects in a phenomenological way.\\ This
indicates that present $3d$ and $2d$ Lagrangian data are {\it
 not inconsistent with} the relation (\ref{bridge}), once finite Reynolds
number effects are kept into account. Furthermore, the use of the
Batchelor parameterization in $3d$ turbulence allows to make
prediction on the values of Reynolds number for which a given window
of direct scaling is expected to appear in the second order moment.

Alternatively, one might not follow the Ockham's suggestion {\it
  ``Entia non sunt multiplicanda praeter necessitatem''} \cite{occam}
and invoke a genuine -i.e. not Reynolds number dependent- departure of
$S_2(\tau)$ from the linear scaling predicted by (\ref{bridge}). For
instance, in \cite{loro} it was investigated the possibility that
anomalous scaling develops already for $S_2(\tau)$ and it was showed
that also this option {\it is not inconsistent with the} data.\\ More
investigation will be needed to understand whether the simple
description (\ref{bridge}), plus Reynolds number effects, is all we
need -as far as scaling properties are concerned-, or if anomalous
scaling as suggested in \cite{loro} is correct.\\
Here, we also showed that (\ref{bridge}) is able to predict the
Reynolds number dependency of the normalized root mean squared acceleration
without the need to introduce any free parameter, if multifractal
fluctuations in the Eulerian statistics are considered. 

Finally, let us comment that in $3d$ turbulence, different scaling
exponents for transverse and longitudinal spatial increments are
observed \cite{gotoh_vel,JFM}, something not fully understood. Along a
Lagrangian trajectory, both longitudinal and transverse Eulerian
fluctuations are naturally mixed and entangled, introducing some
uncertainties in the {\it bridge} relation as discussed here. In the
$2d$ inverse cascade regime, Eulerian longitudinal increments do not
show any deviation from Gaussianity, while the excess kurtosis
measured on a mixed longitudinal and transverse Eulerian increments is
different from zero. The Lagrangian equivalent of the latter Eulerian
measurement is also non Gaussian and in agreement with the {\it
  bridge} relation. Therefore, there are still many open points that
must be further clarified.\\

We acknowledge useful discussions with E. Bodenschatz, M. Cencini,
G. Falkovich, S. Musacchio, A. Pumir, P.K. Yeung and H. Xu. We thank
P.K. Yeung for sharing with us some unpublished data at $Re_\lambda =
1000$ on the acceleration variance, shown in Fig.~4. We thank the
Kavli Institute for Theoretical Physics four our stay in the framework
of the 2011 {\it The Nature of Turbulence} program. This research was
supported in part by the National Science Foundation under Grant
No. PHY05-51164. Also, this research was partly supported by the
Project of Knowledge Innovation Program (PKIP) of Chinese Academy of
Sciences, grant No. KJCX2.YW10 and by the Cost Action MP0806. We thank
Kavli Institute for Theoretical Physics China for our stay in the
framework of the 2012 {\it New Directions in Turbulence} program. We
also acknowledge support from CINECA and Caspur.


\end{document}